# Performance Modelling of Electro-optical Devices for Military Target Acquisition

S.K. Das and R.S. Singh

*Institute for Systems & Studies and Analyses, Delhi-110 054*

**ABSTRACT**

Accurate predictions of electro-optical imager performance are important for defence decision-making. The predictions serve as a guide for system development and are used in war game, simulations that directly influence engagement tactics. In the present study, mathematical models have been developed which involves detection of different military targets using their opto-electronics properties in different environmental conditions. The method first calculates the signal-to-noise ratio received by the observing sensors reflected from the target by quantifying the light energy in terms of photons, which is used for evaluating the detection probability.

## 1. INTRODUCTION

Acquisition of target is a vital task prior to target engagement. The detection or discovery of a target is a classical statistical problem of finding a signal in a background of noise[1]. A signal is a discrete and measurable event produced by a target, whereas noise is any process or phenomenon unrelated to a target that can mask or be mistaken for the target. Target detection systems, regardless of which sensors these are based on, measure a combination of both signal and noise. Reliable detection can only be accomplished when the signal can be clearly distinguished from the noise. Jones[2] assumed signal plus noise and noise follow Gaussian distribution with overlapping mean and standard deviations. Using that Gaussian nature of signal and noise he stated that one can determine the signal-to-noise ratio (SNR) required by an opto-electrical device to detect a signal with a given reliability and a given false alarm rate.

The process of optically detecting a difference in brightness between two adjacent object elements depends on the ability of the sensor to distinguish between the numbers of photons it receives and registers from the two elements discretely[3]. The difference between the two quanta of photons gives rise to a signal. Since the emission of photons is a random process, the statistical fluctuations in these numbers cause an associated noise. Statistical fluctuations in the arrival of photons limit the contrast perception of the eye. Rose[4] made some approaches to the quantitative effect of these fluctuations.

Target acquisition is complex. Many models of the process have been developed, and often these are specialised to only a few military scenarios. For most models only partial validation exists due to the difficulties in carrying out realistic field tests[5].

Recently, Sheffer[6], *et al*. and Birkemark[7] studied the robustness of different statistical criteria such as signal-to-clutter ratio (SCR), Mahalanobis distance, Bhattacharya distance, and Information criteria for target and background seperability in different spectral wavelength of electromagnetic radiation.

To evaluate the target detection capability by various sensors like binocular, image intensifier (II), night vision devices (working in visual wavelength) and thermal imagers (working in the thermal wavelength), some mathematical models have been developed and discussed below.

## 2. APPROACH

Performance of a sensor is defined in terms of the probability of detection, or $P_d$, which is the likelihood that a device will detect a real target of given size at defined range given a probability of false alarm, where $P_{fa}$, which is the likelihood that a device may declare the presence of a target of a given dimension at some range and prevalent environmental conditions falsely.

The matrix in Fig. 1 shows the possible outcomes of a target detection test. When the measurements match actual conditions, the result is a correct test decision, either the detection of an actual target or the confirmation that none exists. If the measurements do not match actual conditions, the test decision is incorrect–either a missed detection or a false alarm.

Let it be assumed that the noise involves a Gaussian distribution of amplitudes. To be sure the distribution of photons is Poisson rather than Gaussian, but except when the number of photons is small, the two distributions are practically indistinguishable. Suppose now that one wishes to detect a signal in the presence of Gaussian noise. The situation is illustrated in Fig. 2. In Fig. 2, the Gaussian curve on the left represents the distribution of amplitudes when the signal is absent, and the similar curve on the right is the distribution of signal-plus-noise amplitudes when the signal is present. If the decision threshold is at the position $T$, then the shaded area to the right of the vertical line at $T$ is the probability $P_{fa}$ that the device falsely concludes that a signal is present when it is not [Type I error, Fig 2(a)], and the area to the left of the vertical line is the probability $1-P_d$ that the device concludes that a signal is not present when actually it is [Type II error, Fig 2(b)].

Figure 2 establishes a graphical relation between the signal-to-noise ratio (SNR) $k$, the false alarm fraction $P_{fa}$, and the detection probability $P_d$. To obtain the corresponding relation, one define $x = erf^{-1}(y)$ as the relation that is inverse to

$$y = erf(x) \equiv \frac{1}{\sqrt{2\delta}} \int_{-\infty}^{x} \exp(\frac{-u^2}{2}) du \quad (1)$$

where $erf(x)$ is the well-known error function. We

**Figure 1. Possible outcomes of a target detection test.**

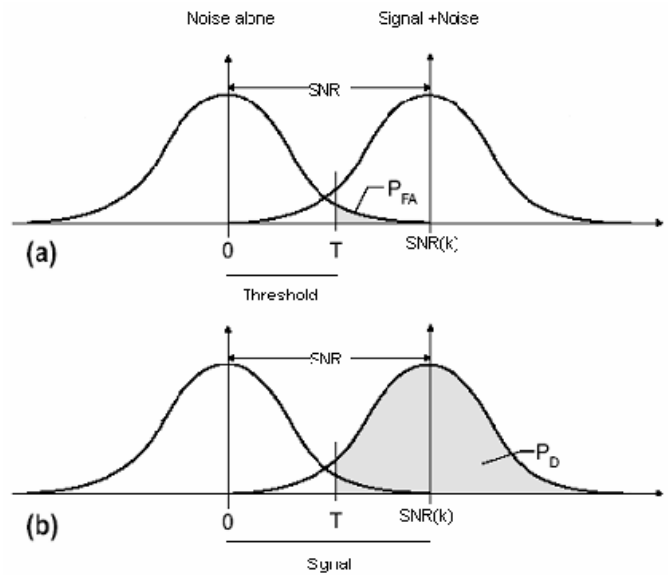

**Figure 2. Determining the SNR: (a) probability of false alarm, and (b) probability of detection.**

suppose, the left Gaussian curve to be expressed by

$$Y=(2\pi)^{-1/2}\exp[-x^2/2] \sim N(0,1) \quad (2)$$

and the right Gaussian curve by

$$Y=(2\pi)^{-1/2}\exp[-(x-k)^2/2] \sim N(k,1) \quad (3)$$

Both Gaussian curves have unit area and unit standard deviation, when $k$ is the ratio of the signal amplitude to the root mean square (rms) noise amplitudes. For an ideal device, if the measured amplitudes is above the threshold, the device concludes that the signal is present, otherwise noise is present.

Then it is easy to show that the SNR, $k$ required by the ideal device of Fig. 1 to achieve a detection probability $P_d$ with a false alarm fraction $P_{fa}$ is derived as

$$1-P_d = \frac{1}{\sqrt{2\delta}} \int_{-\infty}^{T} \exp(\frac{-(u-k)^2}{2}) du \quad (4)$$

Let $u-k = v$, then $du = dv$, When $u \to T$, $v \to T-k$

Thus,

$$1-P_d = \frac{1}{\sqrt{2\delta}} \int_{-\infty}^{T-k} \exp(-\frac{v^2}{2}) dv \quad (5)$$

$$\therefore 1-P_d = erf(T-k) \quad (6)$$

or $P_d = 1-erf(T-k)$ \quad (7)

Similarly from Eqn (1)

$$1-P_{fa} = \frac{1}{\sqrt{2\delta}} \int_{-\infty}^{T} \exp(\frac{-u^2}{2}) du = erf(T) \quad (8)$$

Therefore, $T = erf^{-1}(1-P_{fa})$ \quad (9)

$$\therefore P_d = 1-erf\ [erf^{-1}(1-P_{fa})-SNR] \quad (10)$$

Thus by inverse interpolation of normal tables one can determine the SNR required by an ideal device to detect a signal with a given false alarm fraction $P_{fa}$.

## 3. DETECTION MODEL FOR HUMAN EYE, BINOCULAR, AND IMAGE INTENSIFIER

The basic premise of this model is to determine probability of detection of a target, by a viewing device, in specified environmental conditions, at different ranges. The model is based on the quantum mechanics of light. The model uses the following parameters, in processing the signal, for ascertaining the probability of detection:

(a) Sun or moon illumination ($I$ in lux)
(b) Range, $R$ (km)
(c) Target size, $s$ (cm$^2$)
(d) Target reflectance, $R_t$
(e) Background reflectance, $R_b$
(f) Aperture of the sensor/ pupil radius, $r$ (mm)
(g) Detector efficiency, $\theta^2$
(h) Integration time (eye response time)[4,8] $\tau$, (s)
(i) Photon intensity[8] (i.e., number of photons per lumen per second), P
(j) False alarm rate, $P_{fa}$
(k) Attenuation coefficient, AC

The algorithm for visual detection proceeds as follows:

(i) Based on ambient illumination, target reflectance and background reflectance, the brightness status of the target [$L_t = (R_t * I)/\pi$ in cd/m$^2$] and the background [$L_b = (R_b * I)/\pi$] is evaluated. From these, the contrast ratio [$C=(L_t-L_b)/(L_t+L_b)$] between the target and background is evaluated. Then, using range and attenuation coefficient, apparent contrast [$C'= C*\exp(-AC*R)$] of the target is calculated.

(ii) Depending on the range and target size, the angle ($a = 57.3*60\ s/R$) projected by the target on the detecting device is determined. Using the information on viewing device aperture or pupil diameter and the light energy reflected by target and background, the number of photons in the two energies is calculated. Considering the efficiency of the viewing system, eye response time and photons received, the signal strength can be quantified.

(iii) From signal strength, noise component is identified and SNR is calculated. Based on SNR= $\{2.66*10^{-11}L_m\ 2C^2a^2\ \theta\ P\ r^2\tau\}^{1/2}$ and assumed false alarm rate ($P_{fa}$), probability of detection[3] ($P_d=1-erf\ (erf^{-1}(1-P_{fa})-SNR)$) can be evaluated using the approach outlined in Section 2; where $L_m$ is the mean brightness and $erf$ stands for error function.

## 4. RECOGNITION MODEL FOR THERMAL IMAGER

Detection is less important for thermal acquisition of targets. Temperature difference of targets and background are the important factors for thermal recognition.

The basic premise of this model is to determine probability of recognition of a target, by a thermal imager, in specified environmental condition, at different ranges. The model uses the following parameters, for ascertaining the probability of recognition.

(a) Inherent temperature difference of target and background ($\Delta T_i$)

(b) Range ($R$)

(c) Target height ($H_{targ}$)

(d) Parameters of minimum recognisable temperature difference (MRTD) curve of thermal imager ($a$, $b$). These are the ordinary least square (OLS) estimators (see Appendix I).

(e) Attenuation coefficient ($AC$)

The algorithm for thermal recognition proceeds as follows:

(i) Determine the target critical dimension ($H_{targ}/R$), and apparent temperature difference [$\Delta T_a = \Delta T_i * (-AC*R)$], using knowledge of the atmospheric attenuation coefficient ($AC$) and range ($R$)

(ii) Calculate or measure the system MRTD [$a*\exp(b*SF)$] where SF stands for spatial frequency which is reciprocal of the target critical dimension. From the apparent ($\Delta T_a$) and the MRTD determine the maximum resolvable spatial frequency [$fx=(1/b)*\log(\Delta T_a/a)$] of the sensor

(iii) Using the angular subtends ($f_x$) of the target critical dimension and $H_{targ}/R$, calculate the maximum number of resolvable cycles [$N=f_x*(H_{targ}/R)$] across the target

(iv) Determine the probability of recognition from the target transform probability function (TTPF) curve[5] as

$$P_r = \frac{(N/N_{50})^E}{1+(N/N_{50})^E}$$

where, $P_r$ is probability of recognition; $N$ is maximum resolvable cycles across the target; $N_{50}$ is resolvable cycles for 50 per cent probability of recognition; and $E=2.7+0.7(N/N_{50})$

## 5. RESULTS

In this paper an attempt has been made to evaluate the performance of human eye in different illumination conditions for some targets. The performance is determined in terms of detection probability, which is represented as a function of range. Night-seeing capability of human eye can be enhanced using image intensifier. Figures 3(a)-3(d) show the performance of human eye, image intensifier, and binocular, using the parameters value given in Table 1. The human eye can detect the target according to their size upto 1.5 km (with 100 % detection probability) in daylight without using viewing devices. If viewing devices (binocular) are used, then the range enhances to 5 km under the same weather conditions. Similarly human eye can detect the targets according to their size up to 500 m (with 100 % detection probability) in fullmoon conditions without using any night vision devices. If image intensifier is used, then the range enhances to 2 km under the same weather conditions. Further, it is apparent from the curves that as the range increases, the detection probability decreases sharply and becomes asymptotic around 20 per cent.

A similar exercise has been conducted for target recognition using the thermal imager at night. Figure 4 shows the performance of thermal imager for recognising a tank target using the parameters value given in Table 2. Under clear weather conditions, an average thermal device can recognise a tank at a distance around 3500 m, whereas in poor weather condition, the thermal device can recognise a tank only at a distance around 2000 m.

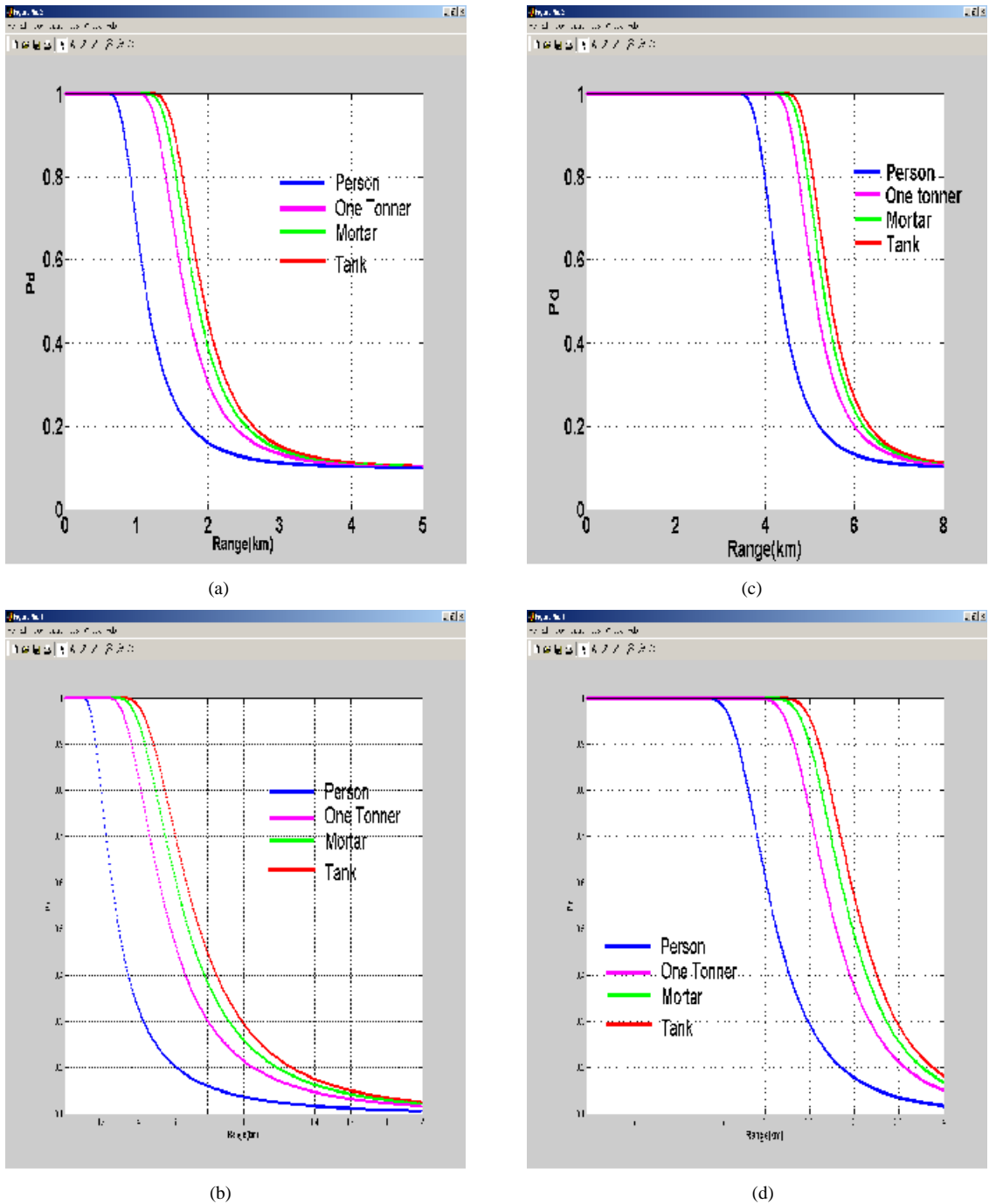

**Figure 3. Detection probability of different targets at clear weather condition, as a function of range (km): (a) daylight with clear weather by an unaided human eye, (b) fullmoon night by an unaided human eye, (c) daylight with binocular, and (d) fullmoon night by an image intensifier-II.**

## 6. CONCLUSIONS

Some of the significant observations are:

(a) Contrast is the only relative measure of target's brightness with uniform background brightness,

**Table 1. Parameters value for evaluating performance of human eye, image intensifier, and binocular for day and night vision.**

| Sl. No. | Parameters | Specifications | Daylight | Fullmoon |
|---|---|---|---|---|
| 1 | Illumination | Day light | 10.0E+2 lux | 3.1E-01 lux |
| 2 | Range | | 0-5 km | 0-2 km |
| 3 | Target size | Person | 1.6 m$^2$ | 1.6 m$^2$ |
| | | Tank | 2.5 m$^2$ | 2.5 m$^2$ |
| | | 1 Toner | 2.0 m$^2$ | 2.0 m$^2$ |
| | | Mortar | 1.4 m$^2$ | 1.4 m$^2$ |
| 4 | Target reflectance | Person | 23.6 % | 23.6 % |
| | | Tank | 20 % | 20 % |
| | | 1 Toner | 19.6 % | 19.6 % |
| | | Mortar | 19 % | 19 % |
| 5 | Background | Desert, reflectance | 26 % | 26 % |
| 6 | Sensors aperture | Pupil radius | 0.1 cm | 0.4 cm |
| | | Binocular/image intensifier (II) NOD MK I | 8.0 cm | 20 cm |
| 7 | Sensors efficiency | Photopic vision/mesophic vision | (1/4200) | (1/4200) |
| 8 | Integration time | -- | 0.1 s | 0.2 s |
| 9 | Photon intensity | Photopic vision/mesophic vision | 4.073E+13 photons/ lumen-s | 4.073E+13 photons/ lumen-s |
| 10 | Probability of false alarm | -- | 1.00E-1 | 1.00E-1 |
| 11 | Attenuation coefficient | Clear environment | 1.118 | 1.071 |

which is used for the visual detection. As SNR is directly proportional to the square of target's contrast, minor changes in contrast lead to significant change in SNR.

(b) The detection probability is highly sensitive to the background illumination. Under night condition such as starlight, human eye can detect a tank at a distance around 200 m, where as at fullmoon condition, the background illumination is high, the human eye in fullmoon illumination can detect a tank at a distance around 1500 m with 50 per cent probability of detection.

(c) The efficiency of photopic vision is maximum at wavelength around 600 nm, at this range, the photon intensity is 4.2 x 10$^{13}$ per lumen-second. Whereas the efficiency of scotopic vision is maximum at wavelength around 500 nm, at this range the photon intensity is 3.5 x 10$^{13}$ per lumen-second. As the SNR is directly proportional to the photon intensity, under these conditions, human eye can detect the same target at longer range in day as compared to at night.

(d) Angle subtended by a target to the detector is an important factor as compared to target's size. Angle is a function of target's size and distance. Different targets can produce same angle at different distances. SNR is directly proportional to the square of target's angle. So little change in target's angle can change the SNR significantly.

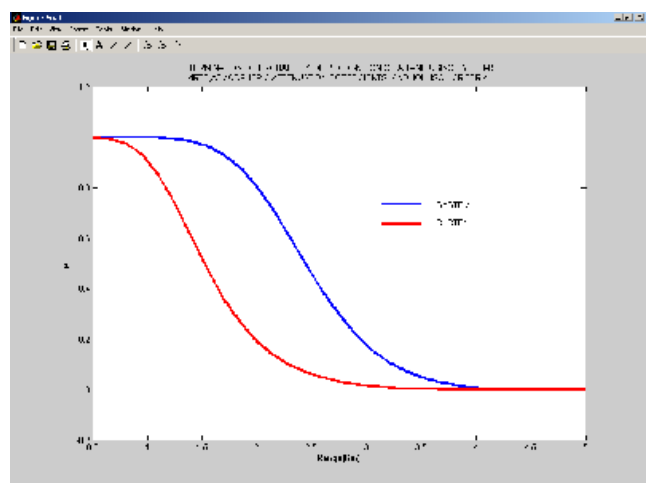

**Figure 4. Probability of recognition of a tank in clear night by Systems I and II thermal imager using the parameter values given in the Table 2.**

**Table 2. Parameters value for evaluating performance of thermal imager**

| Sl No. | Parameters | Specifications | Values |
|---|---|---|---|
| 1. | Inherent temperature difference of target and background | Tank | 1.25 °C |
| 2. | Range | -- | 3 km |
| 3. | Target height | Tank | 2.5 m |
| 4. | Parameters (a, b) of minimum recognisable temperature difference (MRTD) curve of thermal imager | (i) System I<br>(ii) System II | (0.0106, 0.584)<br>(0.0145, 1.1942) |
| 5. | Attenuation coefficient | Clear | 1.071 |

(e) Number of photons captured by the pupil (human eye) depends on its diameter. The pupil diameter of the human eye changes with the background illumination. Number of photons captured by human eye is directly proportional to the square of the pupil diameter. The pupil diameter can be increased using an artificial pupil that will capture maximum number of photons. Using an artificial pupil of 80 mm dia, the detection range of a tank can be increased up to 500 m in starlight condition and 2500 m in full moon condition with 50 per cent detection probability.

(f) Under good weather condition, an average thermal device can recognise a tank at around 3500 m, whereas at poor weather condition, the thermal device can recognise a tank only at around 2000 m.

## ACKNOWLEDGEMENTS

The authors thankfully acknowledge the help of Sh. Rajiv Gupta and Sh. Debashish Dutta, both Scientist F, ISSA, in conducting this research and the constructive comments received from the anonymous referees.

*Appendix* 1

## Parameters of Thermal Imagers

### System I

An optical system using F/2.0, 200 mm focal length and having an overall transmission of 0.5 has been chosen. The detector element (MCT-PV type) chosen is 2 x 15 serial-parallel scan type with a scanning efficiency of 0.66 for a 14-facet opto-mechanical scanner. The elemental size of the detector is 35 µ x 35 µ with a gap of 10 µ between two successive rows and 100 µ between the two columns. The detectivity of the detector has been assumed to be 6 x 1010 cm $Hz^{1/2}W^{-1}$. The scanning frame rate has been taken as 25 cycles/s. Table 3 and Fig. 5 show the observed and predicted minimum recognisable temperature difference (MRTD) for System I.

**Table 3. Minimum recognisable temperature difference data for System I**

| Spatial frequency | Observed MRTD | Predicted MRTD |
|---|---|---|
| 1.0 | 0.047863 | 0.0517 |
| 1.5 | 0.08696 | 0.0950 |
| 2.0 | 0.157993 | 0.1632 |
| 2.5 | 0.287048 | 0.2751 |
| 3.0 | 0.521521 | 0.4672 |
| 3.5 | 0.947521 | 0.8151 |
| 4.0 | 1.721495 | 1.5000 |
| 4.5 | 3.127686 | 3.0400 |
| 5.0 | 5.682512 | 7.4950 |

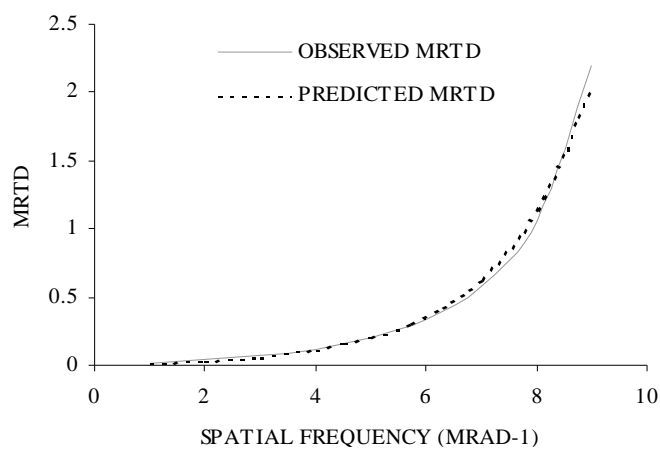

Figure 5. Observed and predicted MRTD for the System I.

### System II

An optical system using F/2.0, 400 mm focal length and having an overall transmission of 0.5 has been chosen. The detector element (MCT-PC type) chosen is 1 x 100 element parallel scan type with 1:2 interlacing and having a scan efficiency of 0.5 for a 12-facet opto-mechanical scanner. The elemental size of the detector is 35 µ x 35 µ with a gap of 35 µ between two successive rows. The detectivity is taken as 8 x $10^{10}$ cm $Hz^{1/2}W^{-1}$ and the scanning frame rate is assumed to be 25 cycles/s. Figure 6 and Table 4 show the observed and predicted minimum recognisable temperature difference (MRTD) for System I.

**Table 4. Minimum recognisable temperature difference data for System II**

| Spatial frequency | Observed MRTD | Predicted MRTD |
| --- | --- | --- |
| 1 | 0.01612 | 0.019008 |
| 2 | 0.03740 | 0.034085 |
| 3 | 0.06878 | 0.061121 |
| 4 | 0.11780 | 0.109602 |
| 5 | 0.19810 | 0.196538 |
| 6 | 0.33490 | 0.352431 |
| 7 | 0.58350 | 0.631978 |
| 8 | 1.06800 | 1.133260 |
| 9 | 2.20000 | 2.032159 |

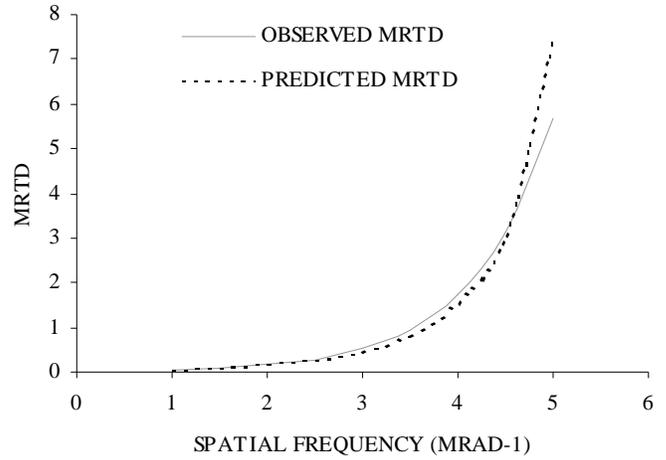

Figure 6. Observed and predicted MRTD for the System II.

## Contributors

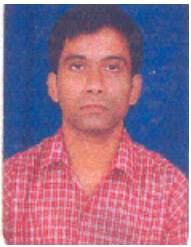

**Dr S.K. Das** obtained PhD (Agricultural Statistics) from the Indian Agricultural Research Institute, New Delhi, in 2005. He joined DRDO at Institute for Systems Studies and Analyses (ISSA), Delhi, in 2002. His areas of research include: Mathematical modelling for military target detection, armour warfare, and digitised battlefield.

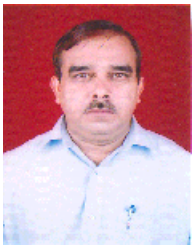

**Mr R.S. Singh** obtained MSc (Mathematics) in 1977. He joined DRDO at the ISSA, Delhi, in 1984. He has worked for reliability evaluation of various systems designed and developed by DRDO systems labs. Presently, he is working for mathematical models development.